\documentclass[pre,twocolumn]{revtex4}
\usepackage{psfrag}
\usepackage{amssymb,amsmath,amsthm}
\usepackage[dvips]{graphicx}
\usepackage{verbatim}
\usepackage{color}
\usepackage{amsmath}
\usepackage{latexsym}
\usepackage{amsfonts}
\usepackage{amssymb}
\usepackage{txfonts}
\usepackage{pxfonts}
\usepackage{wasysym}

\definecolor{amber}{rgb}{1.0, 0.49, 0.0}

\begin{document}
\title{Analysis of dysautonomia in patients with Chagas Cardiomyopathy}

%\footnote{supported by Universidad Nacional de San Agust\'in de Arequipa, Per\'u under the grant 01-2018-UNSA }
\author{ M. Vizcardo$^{\dag}$, A. Ravelo$^{\ddag}$,  P. Gomis$^{\S}$}
%\author{ M. Vizcardo$^{\dag}$, A. Ravelo$^{\ddag}$}

\affiliation{$^{\dag}$Departamento Acad\'emico de F\'isica, Universidad Nacional de San Agust\'in de Arequipa, Av. Independencia, Arequipa, Per\'u,mvizcardoc@unsa.edu.pe\\
$^{\ddag}$Instituto for Technological Develoment and Innovation in Communications, Universidad Las Palmas de Gran Canaria, Spain\\
$^{\S}$ Valencian International University, Spain}

\begin{abstract}
\noindent 
Chagas disease American trypanosomiasis is caused by a flagellated parasite: trypanosoma cruzi, transmitted by an insect of the genus Triatoma and also by blood transfusions. In Latin America the number of infected people is approximately 6 million, with a population exposed to the risk of infection of 550000. It is our interest to develop a non-invasive, low-cost methodology, capable of detecting any alteration early on cardiaca produced by T. cruzi.
We analyzed the 24 hour RR records in patients with ECG abnormalities (CH2), patients without ECG alterations (CH1) who had positive serological findings for Chagas disease and healthy (Control) matched by sex and age.
We found significant differences between the Control, CH1 and CH2 groups that show dysautonomy and enervation of the autonomic nervous system.

\bigskip


\end{abstract}

\maketitle

\noindent {\bf 1. Introduction} \smallskip\\

Chagas disease is caused by a flagellated parasite: Trypanosoma cruzi, transmitted by an insect of the genus Triatoma and also by blood transfusions. In Latin America the number of people infected is approximately 6 million, with a population exposed to the risk of infection of 568,000\cite{med:oms2017}. In 40 \% of the population infected with Trypanosoma cruzi (T. cruzi) there is cardiac involvement \cite{med:fm78, med:fm80, med:hag91}. These estimates explain why this disease is a serious public health problem in the countries where it is endemic. In the evolution of Chagas disease we can distinguish an initial acute phase of infection and a prolonged intermediate chronic phase, in which the disease is often clinically silent, and the usual diagnostic techniques do not provide a robust criterion to predict whether a seropositive asymptomatic patient will suffer cardiac involvement. 
In the last two decades we have known the relationship between the autonomic nervous system and cardiovascular mortality, including sudden death.
Experimental evidences of the association between the probability of having malignant arrhythmias and the increase of the sympathetic tone or the decrease of the vagal tone, have stimulated the development of quantitative markers of the activity of the autonomic nervous system.
The variability of the heart rate represents one of the most promising markers. The term "Heart Rate Variability (HVR)" defines the variations in the intervals between the consecutive heartbeats or consecutive R-R interval of electricity. It is used to describe changes in the duration of cardiac cycles, as well as instantaneous variations in heart rate. During the last twenty years a close association has been observed between the functioning of the autonomic nervous system and mortality due to cardiovascular causes. The experimental evidence that the increase in sympathetic activity or the reduction of parasympathetic disease predict lethal arrhythmias, has generated the intense search for a method that quantifies the influence of the autonomic nervous system on the heart.

It is our interest to develop a non-invasive low-cost methodology, that allows to see the dysautonomia or dysfunction in the course ofthe 24 hours. We hypothesize that some kind of dysautonomia measured with HRV tecniques can early detect any cardiac alterations produced by the T. cruzi.
\noindent 

\bigskip

\noindent {\bf 2. Database and Registry}\smallskip\\ 

The results of this work were obtained by processing the electrocardiogram (ECG) and obtaining the RR interval, from three different groups of volunteers. The following test: clinical evaluation, serological Machado-Gerreiro test, chest x-rays, echocardiogram, electrocardiogram and ambulatory Holter registration (24 hours), classify volunteers into three groups: 83 healthy people called {\bf Control} group; 102 patients infected with only positive serology (clinical evaluation, chest x-rays, echocardiogram, electrocardiogram and Holter were normal) called {\bf CH1} group; and 107 seropositive patients with incipient cardiac involvement first degree atrioventricular block (BAV), sinus bradycardia (BS) and or right bundle branch block of the bundle of His (BRDHH),  that were not being treated with medications, called the {\bf CH2} group. All were outpatients, and informed consent was obtained from all of them. The ECG signals were recorded at 500 Hz with 12 bits of resolution, a set of 288 framer of 5 minutes was obtained.

\noindent 

\bigskip

\noindent {\bf 3. Methods} \smallskip \\

We have used the database of the Instituto de Medicina Tropical (IMT) of the Universidad Central de Venezuela. For the detection of the QRS complex the program based on the Pan-Tomkip algorithm\cite{med:pan85} was used, then the 288 tachograms of the RRs of 5 minutes were generated, they were made a post processed the "ada" filter\cite{med:wessel94}, to eliminate the eptopic beat, and then 8 HRV indices were obtained: average RR intervals, standard deviation, pNN50, r-MSSD, LF, HF, LF$/$HF, VLF\cite{med:task96}.\\

For the analysis of the 8 HRV indices we use the PCA Principal Component Analysis method\cite{med:scholkopf} that is to say, reduce the dimension and have an uncorrelated representation. We will mention some characteristics:
\begin{itemize}
\item
Intuitively, the technique serves to determine the number of underlying explanatory factors after a set of data explaining the variability of said data.
\item
The PCA seeks the projection according to which the data is best represented in terms of least squares.
\item
PCA involves the calculation of the decomposition in eigenvalues of the covariance matrix, normally after focusing the data on the average of each attribute.
\item
The calculation of the covariance matrix is based on the internal or scalar product.
\item
Limitations: only linearly separable problems
\end{itemize}
\noindent
The standard linear PCA algorithm:
\begin{itemize}
\item
Given a set of observations $x_i \hspace{0.2cm} \epsilon \hspace{0.2cm}  \mathbb{R}^N \hspace{0.2cm} i= 1,...m$ which are centered, $\Sigma x_i=0$  , The PCA finds the main axes through the diagonalization of the covariance matrix:  $C=\frac{1}{m} \Sigma x_j {x_j}^T$ 
\item
$C$ it is a definite positive matrix, the result of the diagonalization leads to eigenvalues
non-negative To perform the calculation, the equation of eigenvalues is solved: $\lambda \nu=C \nu$ for non-negative eigenvalues and non-zero eigenvectors $\lambda \ge 0$ . 
Replacing :  $\lambda \nu=C \nu =\frac{1}{m} \Sigma <x_j \centerdot {\nu}>x_j$ 
\item
Where you can see that all the solutions $\nu$ con $\lambda \ne 0$ they are in the sub-space generated by $x_1 . . . x_m$, therefore for these solutions the equation of
eigenvalues is equivalent to $\lambda<x_i \centerdot C\nu>$   $\forall i=1,...,m$:
\end{itemize}

\noindent {\bf 4. Results }\\ \smallskip

\noindent {\bf 4.1 Estimators Statistics and Frequency}\\ 

\smallskip
\begin{figure}[!ht]
\includegraphics[scale=0.65,clip=true,angle=0]{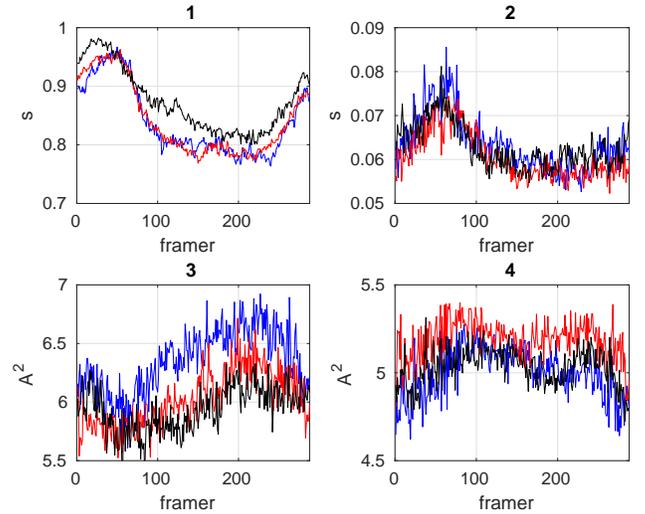}
\label{HRV}\caption{Temporary and Frequency circadian profiles of the average values of: 1) RR intervals, 2) standard deviation, 3) low frequency LF, 4) very low frequency VLF. Groups: Control (blue), CH1 (red) and CH2 (black)}
\end{figure}
With the 288 5-minute tacograms representing 24 hours for each volunteer or patient, 4 statistical indices were obtained (average of the RR intervals, standard deviation, pNN50 and r-MSDD) and 4 spectral indices (LF, HF, LF/HF and VLF). In Figure 1.1, we observe that the average values ​​of the circadian profiles of the CH2 group are above those of the other 2 groups: Control and CH1, this is because the CH2 group is in the dilation phase. With respect to the average values ​​of the circadian profiles: standard deviation (Fig. 1.2), LF (Fig. 1.3) and VLF (Fig. 1.4) we do not see large differences and rather much noise.\\
\begin{figure}[!ht]
\includegraphics[scale=1.2,clip=true,angle=0]{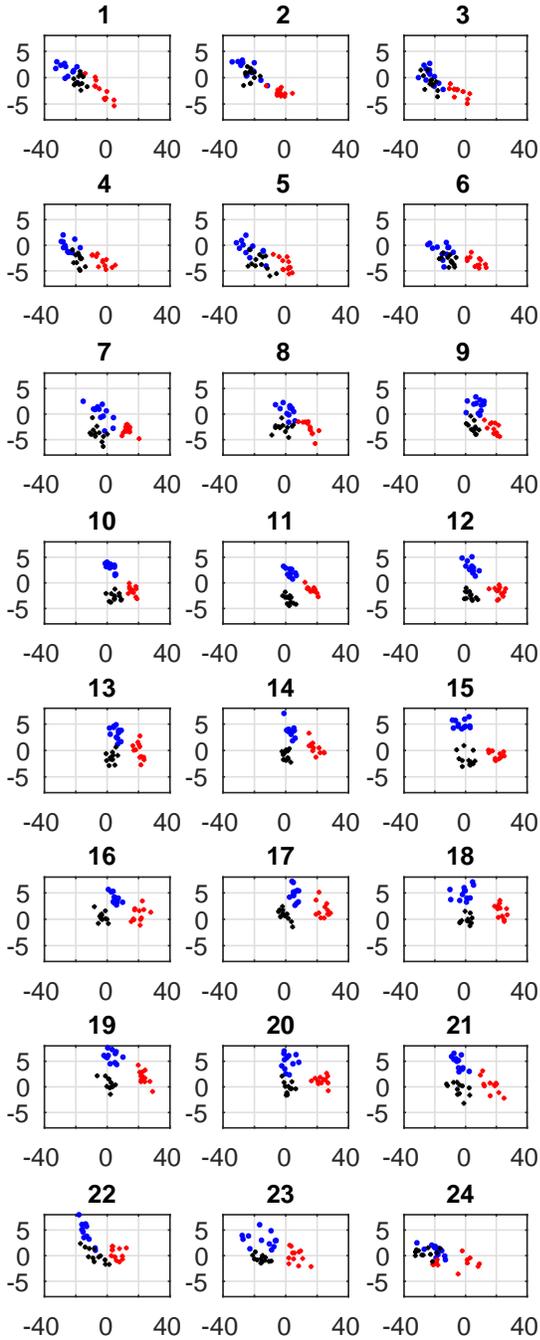}
\caption{First principal components vs second principal components of the 24 hours. Groups: Control (blue), CH1 (red) and CH2 (black)}
\end{figure}

\noindent {\bf 4.2 PCA analysis}\\ 
\smallskip

For the PCA analysis we use the average circadian profiles and the 8 HRV indices as feature of the three groups: Control, CH1 and CH2. The covariance matrix was calculated, then its eigenvalues ​​which were ordered from greater to less with their respective eigenvectors, the first  and second eigenvalue will have the greatest variance, that is to rotate and translate two orthogonal axes in the direction of greater variance. These axes are called principal components (PC).\\

On the first principal component (1st PC) and second principal component (2nd PC) we project the 12 vectors corresponding to each hour of the three groups and are shown in Fig. 2, we do this for each hour of the 24 hours.\\

We calculate the centroid of each group in each hour and see its evolution in 24 hours (Fig. 3) that would be the descriptors of the dynamics of each group.\\

We have identified three stages in the dynamic: from 01 to 09 hours, from 09 to 20 hours and from 20 to 24 hours.\\

\begin{figure}[!ht]
\includegraphics[scale=0.61,clip=true,angle=0]{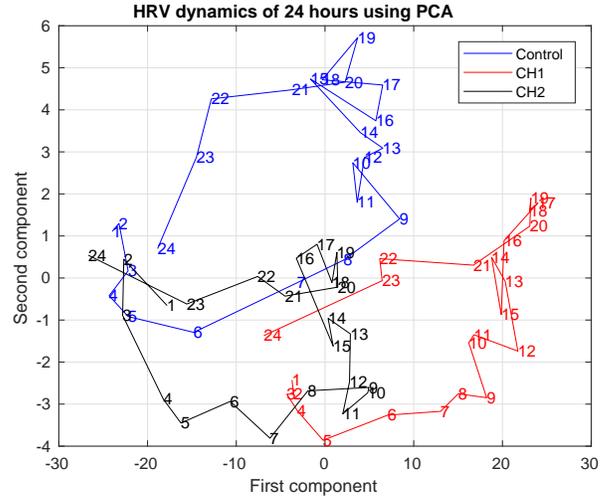}
\caption{First  component principal versus second principal component of the 24 hour average hourly values. Groups: Control (blue), CH1 (red) and CH2 (black)}
\end{figure}

\noindent {\bf 4.2.1  From 01 to 09 hours}\\

Linear regression was made between 01 and 05 hours (Fig. 4) to the three groups:\\
Control $y_{1-5}=-0.31x-6.9$\\
CH1 $y_{1-5}=-0.31x-3.9$\\
CH2 $y_{1-5}=-0.41x-9.5$\\
Then between 05 to 09 hours:\\
Control $y_{5-9}=0.08x+0.38$\\
CH1 $y_{5-9}=0.057x-3.8$\\
CH2 $y_{5-9}=0.038x-2.9$\\

\begin{figure}[!ht]
\includegraphics[scale=0.61,clip=true,angle=0]{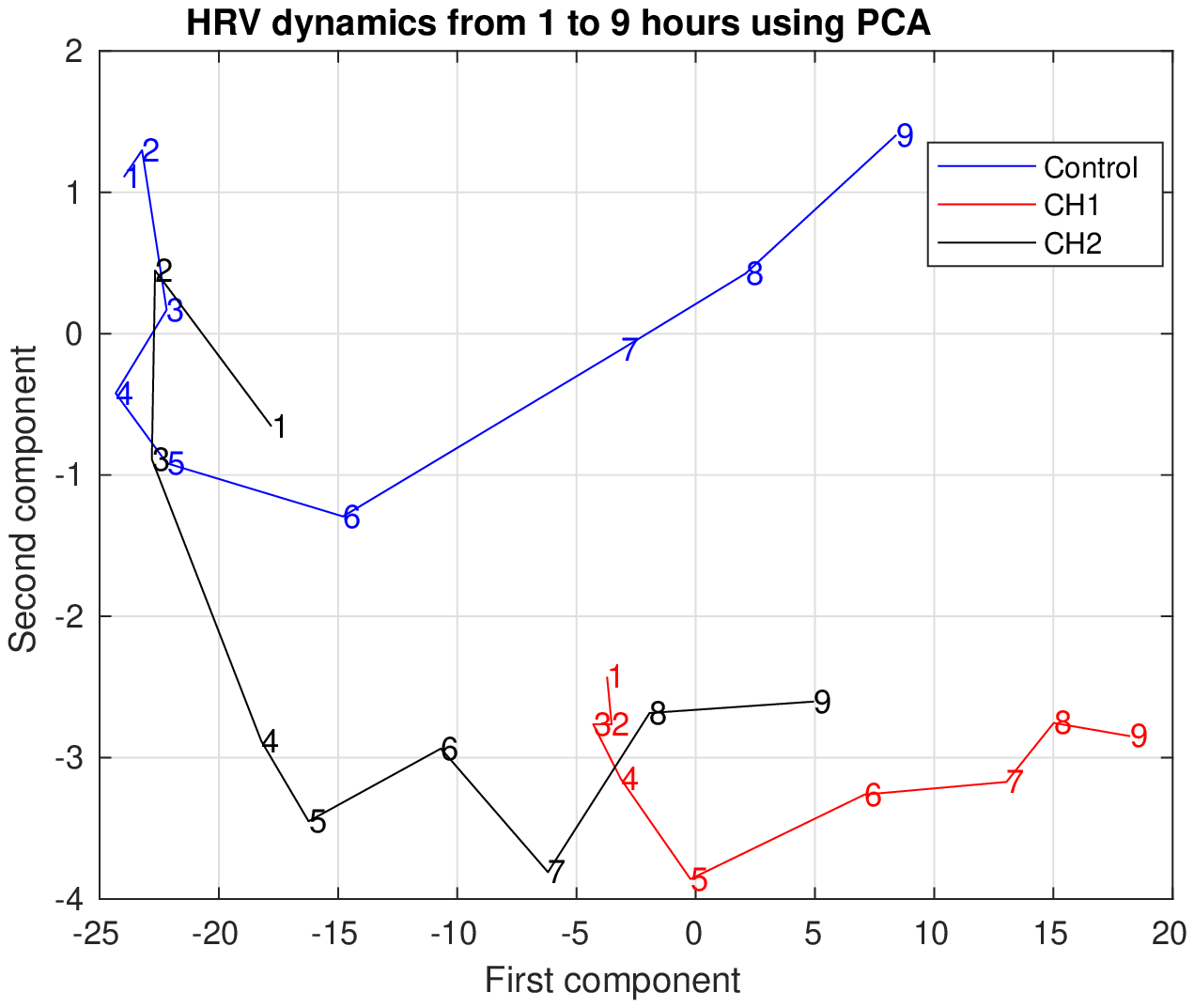}
\caption{First principal components vs second principal components of the 24 hours, groups: Control (blue), CH1 (red) and CH2 (black)}
\end{figure}

\noindent {\bf 4.2.2 From  09 to 20 hours}\\

 We observe that the dynamics of each group is well defined (Fig 5) and that the linear regression of the three groups has been calculated:\\
Control $y_{9-20}=-0.23x+4.5$\\
CH1 $y_{9-20}=0.45x-9.3$\\
CH2 $y_{9-20}=-0.47x-0.41$\\

\begin{figure}[ht]
\includegraphics[scale=0.61,clip=true,angle=0]{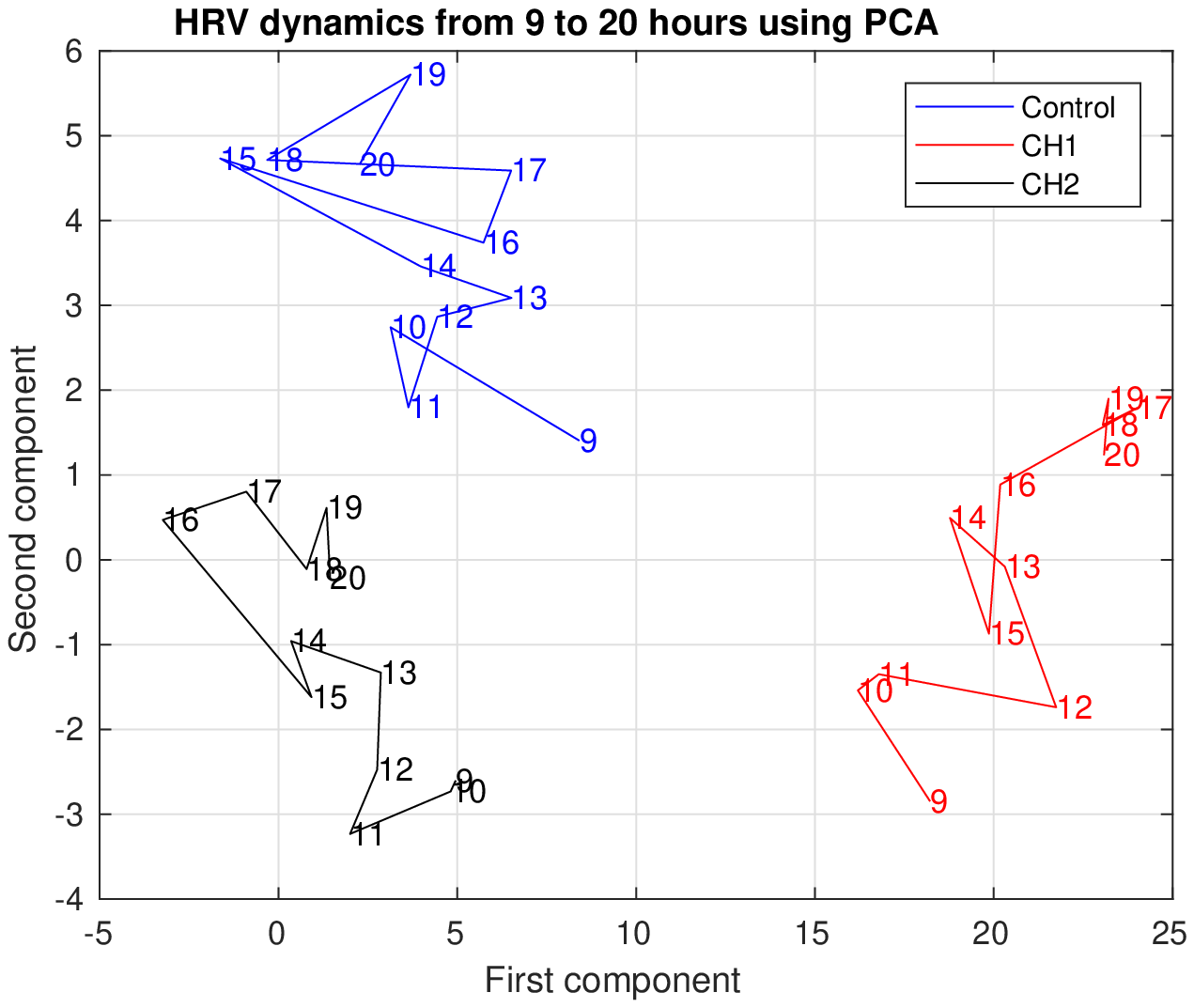}
\caption{First principal components vs second principal components of the 24 hours, groups: Control (blue), CH1 (red) and CH2 (black)}
\end{figure}
\noindent {\bf 4.2.3 From 20 to 24 hours}\\

\begin{figure}[!ht]
\includegraphics[scale=0.61,clip=true,angle=0]{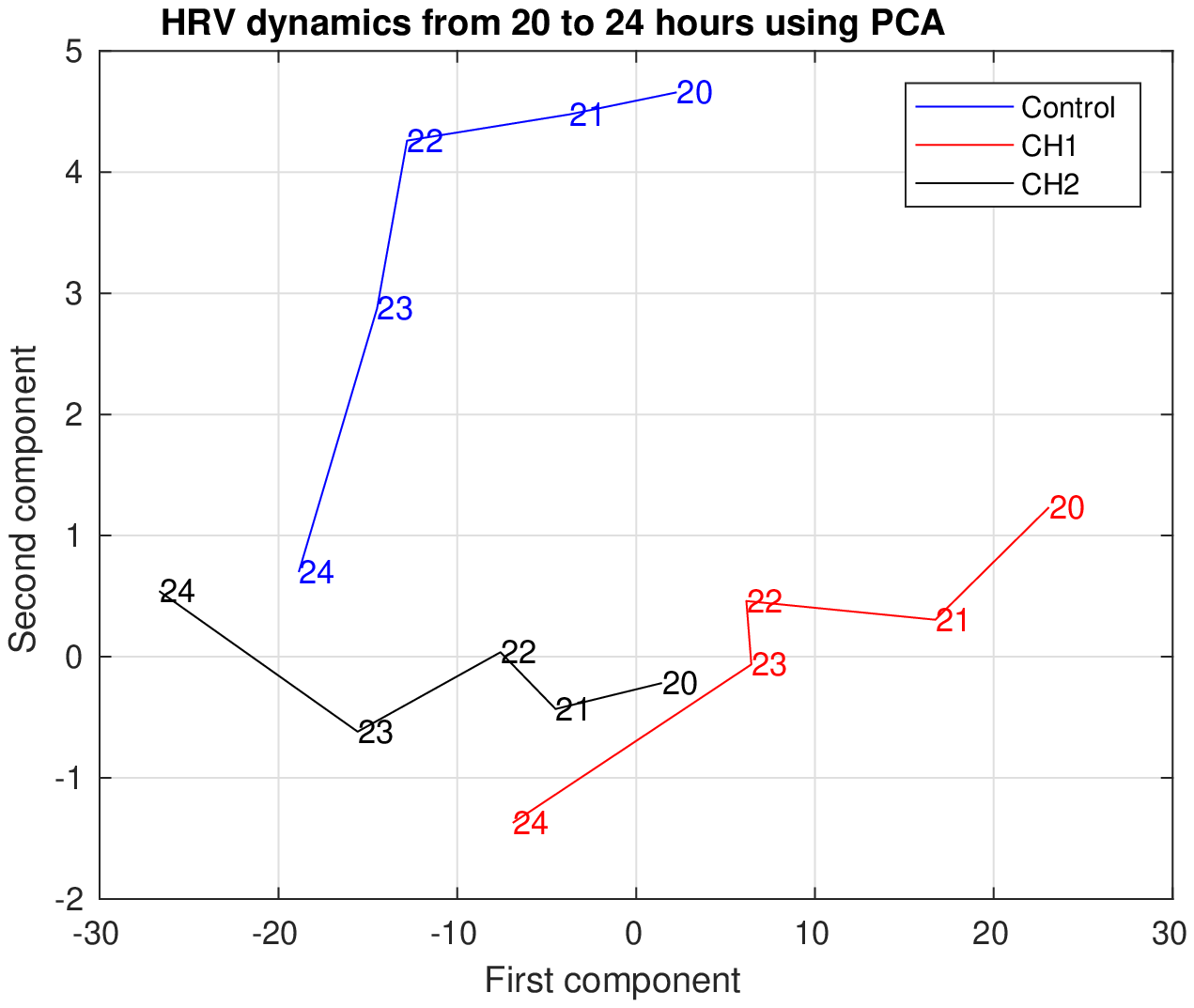}
\caption{First principal components vs second principal components of the 24 hours, groups: Control (blue), CH1 (red) and CH2 (black)}
\end{figure}
The linear regression of the three groups (Fig. 6) was calculated:\\
Control $y_{20-24}=0.16x+4.9$\\
CH1 $y_{20-24}=0.077x-0.59$\\
CH2 $y_{20-24}=-0.023x-0.38$\\
\\
\noindent {\bf 4.3 Test Kruskal Wallis at the PCA}\\ 
\bigskip
\begin{table}[htbp]
\begin{center}
\begin{tabular}{|c||c|c||c|c||c|c||}
\hline 
&\scriptsize 1 CP &\scriptsize 2 CP &\scriptsize 1 CP &\scriptsize 2 CP&\scriptsize 1 CP&\scriptsize 2 CP\\ \hline
\scriptsize Hours&\scriptsize C-CH1 &\scriptsize C-CH1&\scriptsize C-CH2&\scriptsize C-CH2&\scriptsize CH1-CH2&\scriptsize CH1-CH2\\
\hline
1& 0.001& 0.001& NS & 0.040& 0.001& NS\\
2& 0.001& 0.001& NS & X & 0.001& 0.001\\
3& 0.001& 0.001& NS & X & 0.001& 0.026\\
4& 0.001& 0.001& NS & 0.001& 0.001& NS\\
5& 0.001& 0.001& NS & 0.006& 0.002& NS\\
6& 0.001& 0.001& NS & 0.027& 0.001& NS\\
7& 0.001& 0.001& NS & 0.001& 0.001& NS\\
8& 0.001& 0.001& NS & 0.001& 0.001& NS\\
9& 0.003& 0.001& NS & 0.001& 0.001& NS\\
10& 0.001& 0.001& NS & 0.001& 0.001& NS\\
11& 0.001& 0.012& NS & 0.001& 0.001& 0.022\\
12& 0.001& 0.001& NS & 0.001& 0.001& NS\\
13& 0.004& 0.004& NS & 0.001& 0.001& NS\\
14& 0.006& 0.006& NS & 0.001& 0.001& NS\\
15& 0.001& 0.001& NS & 0.001& 0.001& NS\\
16& 0.015& 0.001& 0.015& 0.001& 0.001& NS\\
17& 0.013& 0.004& 0.018& 0.001& 0.001& NS\\
18& 0.001& 0.008& NS & 0.001& 0.001& NS\\
19& 0.001& 0.002& NS & 0.001& 0.001& NS\\
20& 0.001& 0.005& NS & 0.001& 0.001& NS\\
21& 0.001& 0.001& NS & 0.001& 0.001& NS\\
22& 0.001& 0.002& NS & 0.001& 0.002& NS\\
23& 0.001& 0.001& NS & 0.001& 0.001& NS\\
24& NS & 0.001& 0.033& NS & 0.001& 0.002\\
%}
\hline
\end{tabular}
\caption{p-value $<$ 0.05 the Kruskal Wallis test }
\label{p-value Kruskal Wallis test  }
\end{center}
\end{table}

The Kruskal Wallis test was used to find significant differences (p-value$<$0.05) between the groups: Control-CH1, Control-CH2 and CH1-CH2 within 24 hours.\\

For this, it was carried out between the 1st PC of the three groups in each hour, then it was done in the 2nd PC, Table 1 shows the p-values ​​less than $ <$ 0.05 and with an "NS" where the values ​​are greater. It can be seen that it is sufficient that one of the main components has $ <$ 0.05 to be significantly different, as can be seen between the Control-CH2 groups in 4 to 23 hours in Fig. 2 and Table 1. Of the 144 p-value there are only 4 that correspond to the hours of 2 and 3 Control-CH2 that are not significantly different.

\normalsize
\noindent {\bf 5. Discussion and conclusions} \\ 
\smallskip

The circadian profiles of the HRV indices do not show significant differences, this may be due to the noise present in the ECGs.\\

The use of PCA with the HRV index achieves significant differences, except in two hours of Control-CH2. \\

We have identified three stages in the 24-hour dynamic; the first where there is a change to the right of the three groups. a second has short displacements and in a defined area and a third where the three groups move to the left. The displacements to the right and to the left are due to the hours of rest (parasympathetic), while the zone of short displacements to the hours of activity (sympathetic)\\

The PCA can be used to obtain HRV dynamics descriptors in 24 hours.\\

We note that in the slope of 1 to 5 hours and from 20 to 24 hours a decrease between the Control Groups, CH1 and CH2, this could use me as a measure of variability.\\

\bigskip

\noindent {\bf Acknowledgments}\\ \smallskip

Universidad Nacional de San Agustin de Arequipa, grant Nro. 01-2018-UNSA \\
Instituto for Technological Develoment and Innovation in Communications, Universidad Las Palmas de Gran Canaria

%Postgrado en Instrumentaci\'on, Facultad de Ciencias, Universidad Central de Venezuela by supporting partial \\
%Consejo Nacional de Investigaciones Cient\'{\i}ficas y Tecnol\'ogicas, CONICIT under the grant G-97000675 by supporting partial \\
\bigskip
\noindent {\bf Conflicts of Interest}\\ \smallskip
The authors declare no conflict of interest\\  
\bigskip

\noindent {\bf References} \\ \smallskip

\bibliography{med}

\end{document}